\documentclass{ragtime} 

\title[Numerical calculations of orbital and epicyclic frequencies of compact stars]%
{Unusual behavior of epicyclic frequencies around rapidly rotating
compact stars}

\author[M. Wi\'sniewicz, 
        D. Gondek-Rosi\'nska,    
        W. Klu\'zniak
        and N. Stergioulas]
       {Mateusz Wi\'sniewicz\at{1,a} 
        Dorota Gondek-Rosi\'nska\at[]{1}, 
        \splitauthors
        W\l{}odzimierz Klu\'zniak\at[]{2}
        and Nikolaos Stergioulas\at[]{3}\\
        \ins{1}Institute of Astronomy, Faculty of Physics \& Astronomy,
        University of Zielona G\'ora,\splitins[1]
        Szafrana~2, 65-001 Zielona G\'ora,
        Poland\\
        \ins{2}Nicolaus Copernicus Astronomical Center, Bartycka 18,
               00-716 Warszawa,
        Poland\\
        \ins{3}Aristotle University of Thessaloniki, Thessaloniki, Greece\\
        \ins{a}\Email{mateusz@astro.ia.uz.zgora.pl}}

\coentry{Z. Stuchl\'{\i}k, J. Kov\'a\v{r} and J. Cimrman}%
       {Equilibrium of spinning test particles in equatorial plane\par
        of Kerr--de~Sitter spacetimes}

\begin{document}
\nocite{*}
\begin{abstract}
We report on numerical calculations of orbital and epicyclic
frequencies in nearly circular orbits around rotating neutron stars
and strange quark stars. The FPS equation of state was used to
describe the  neutron star structure while the MIT bag model was used
to model the equation of state of strange quark stars. The uniformly
rotating stellar configurations were computed in full general
relativity. We find that the vertical epicyclic frequency is very
sensitive to the oblateness of the rotating star.  For models of
rotating neutron stars of moderate mass, as well as for strange quark
star models, the sense of the nodal precession of test particle orbits
close to the star changes at a certain stellar rotation
rate. These findings may have implications for models of kHz QPOs.
\end{abstract}



\section{Introduction}
\label{intro}
The discovery of kHz Quasi-periodic oscillations (QPOs) is among the
most important scientific result  of Rossi X-ray Timing Explorer
(RXTE).  To date, kHz QPOs  have been discovered in about 20 neutron
star low-mass X-ray binaries (LMXBs), which typically exhibit two high
frequency peaks in the power spectra of the X-ray flux.
The QPO phenomenon promises to be a probe of the innermost regions of
accretion disks around compact objects such as white dwarfs, neutron
stars and black holes, although the promise has not yet been fully
realized \citep[see][, for a review]{2000ARA&A..38..717V}. Most models
of  kHz QPOs involve orbital and epicyclic frequencies
\citep{1980PASJ...32..377K,1991ApJ...378..656N,1992ApJ...393..697N,1997ApJ...476..589P,1999ApL&C..38...57S,1999PhR...311..259W,2001A&A...374L..19A,2001ApJ...548..335S,2005AN....326..820K,2012AcA....62..389S,2013A&A...552A..10S}.

In the case of Newtonian gravity of a spherically symmetric body
the three basic frequencies associated
with nearly circular motion, i.e., the orbital ($\Omega_{\rm K}$), radial epicyclic ($\omega_r$), and vertical epicyclic ($\omega_z$) frequencies are equal to each other. In the case of black holes, the radial epicyclic frequency is lower than the orbital one, both in the Schwarzschild and
the Kerr metrics. For prograde orbits of the Kerr black hole,  the vertical epicyclic frequency is lower than the orbital frequency, but higher than the radial epicyclic one, while for retrograde orbits
the vertical epicyclic is larger than the orbital frequency
\citep{1997ApJ...476..589P}.

However, in Newtonian gravity the degeneracy between the three frequencies
can be broken by rotation. It has been shown
that for extremely oblate bodies  in strictly Newtonian gravity
the radial epicyclic frequency
may even go to zero at a certain distance from the body (and be imaginary
closer to it), so that no stable orbits will exist close to a very rapidly
rotating
fluid configuration \citep{2001ESASP.459..301K,2001PhRvD..63h7501Z}.
In particular, no stable circular orbits exist
right outside  Maclaurin spheroids of ellipticity
$e>$0.834583178
\citep{2002A&A...381L..21A}.
Further, it has been shown that in Newton's gravity,
the ordering of the frequencies around an oblate body such as a
Maclaurin spheroid is  $\omega_r<\Omega_{\rm K}<\omega_z$
\citep{2013MNRAS.434.2825K}.
In summary, the Newtonian effects of oblateness are 
the opposite of those of frame-dragging in Kerr geometry 
for prograde orbits:
 the innermost stable circular
orbit is pushed away from the gravitating body and the
vertical epicyclic frequency is increased.

For rotating bodies in general relativity (GR) there is a competition
between frame-dragging and effects of oblateness
\citep{1999A&A...352L.116S}.  Effects of oblateness of the gravitating
rotating body have also been noted in GR in the context of the
``relativistic precession model'' of neutron star kHz QPOs,  which
relies on the differences between the orbital and epicyclic
frequencies \citep{1999ApJ...513..827M}.  In the case of rapidly
rotating strange quark stars \citet{2014PhRvD..89j4001G}, computed the
frequencies for two stellar masses ($M=1.4M_\odot$ and
$M=1.96M_\odot$), and showed that the vertical epicyclic frequency and
the related nodal precession rate of inclined orbits are very
sensitive to the oblateness of the rotating star. In particular, for
rotating stellar models of moderate and high-mass strange quark stars,
the sense of the nodal precession 
(given by the sign of $\Omega_{\rm K}-\omega_z$) 
changes at a certain rotation rate.  We defer a
discussion of the potential astrophysical implications of this finding
till Section \ref{disc}.

We report on numerical calculations of orbital and epicyclic
frequencies for rotating strange quark stars and neutron stars for a
wide range of masses and two rotation rates. We have used the RNS code
for our calculations \citep{1995ApJ...444..306S}. In this contribution
we are discussing the similarities between the behavior of epicyclic
frequencies in strange stars and in neutron stars modeled with the FPS
equation of state.

\begin{figure}
\vspace{4mm}
\begin{center}
\includegraphics[width=0.45\linewidth,height=.35\linewidth]{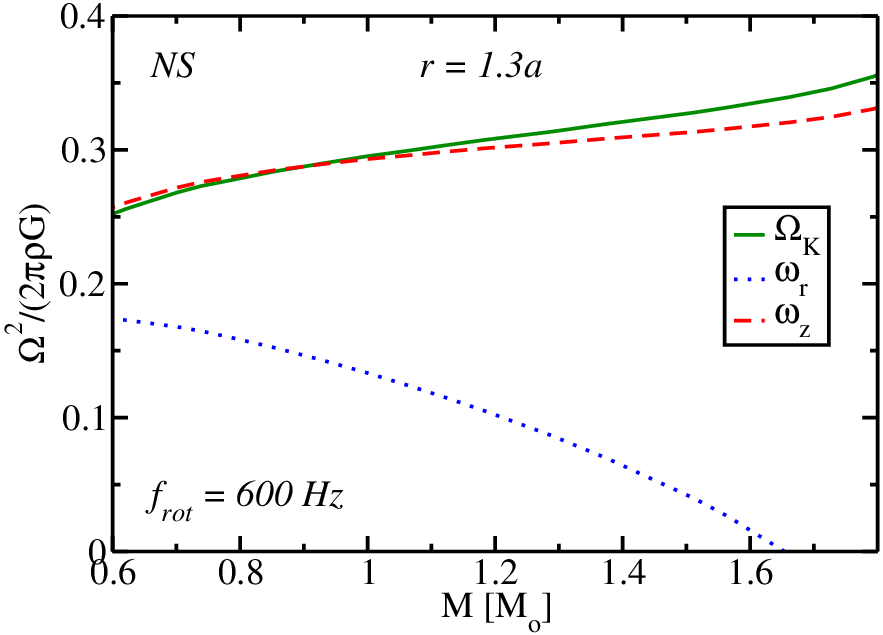}
\hspace{5mm}
\includegraphics[width=0.45\linewidth,height=.35\linewidth]{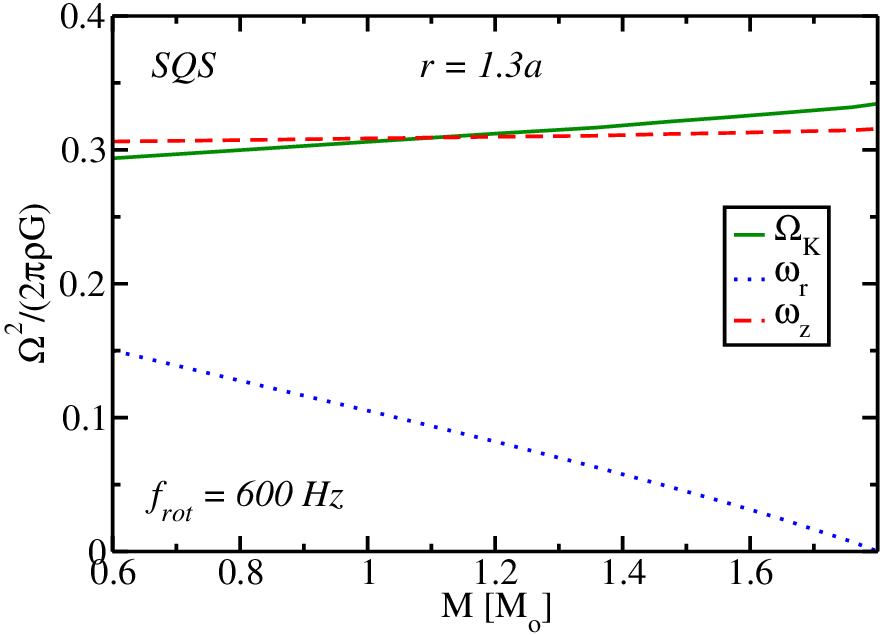}
\end{center}
\caption{Squares of the orbital and epicyclic frequencies (scaled with $2\pi G \rho$) at $r = 1.3a$ versus gravitational mass for uniformly rotating neutron stars (left panel), and strange quark stars (right panel), rotating at a fixed frequency $f_{\rm rot}$ = 600~Hz. The solid (green) line corresponds to orbital frequency, the dashed (red) line to vertical epicyclic frequency and the dotted (blue) line to radial epicyclic frequency.}
\label{fig:one}
\end{figure}

\section{Properties of circular orbits in the metric of rotating 
neutron stars and quark stars}
\label{eom}

We have performed all of our numerical calculations of strange quark
stars in the framework of the MIT bag model
\citep{1984PhRvD..30.2379F}. In this model quark matter is composed of
massless up and down quarks, massive strange quarks and electrons. We
use the simplest MIT bag model with masless strange quarks, with the
equation of state given by
$P = A(\rho - \rho_0)c^2$,
where $P$ is the pressure, $\rho$ is mass-energy density, and $c$ is
the speed of light. We take $A = 1 / 3$ and 
$\rho_0 = 4.2785 \times 10^{14}$~g/cm$^3$.\\
 For computing neutron star models, and their exterior metric, we have used a
modern version of FPS equation of state \citep[see][, for a
  review]{1994ApJ...422..227C} proposed by
\cite{1981NuPhA.361..502F}.  The epicyclic frequencies are computed
from the metric coefficients.
The output of the code includes stellar parameters, such as 
 the gravitational mass $M$, and the
radius of the equator $a$. We find it convenient to determine radial
positions in units of $a$.

 The orbital and epicyclic frequencies are exhibited for a wide range
 of masses of strange stars and neutron stars, for two stellar
 rotation rates, 600~Hz and 900~Hz. In the former case (Fig.~1) we
 present the frequencies squared for orbits at some distance from the
 star (at $r=1.3a$), while in the latter (Fig.~2)  the same quantities
 are displayed for orbits grazing the stellar equator ($r=a$).

The effects of oblateness on the epicyclic frequencies in numerical
solutions for a neutron star rotating at 400 Hz has been clearly seen
in the unusually small difference  $\Omega_{\rm K}-\omega_z>0$ between
the orbital frequency and the vertical epicyclic one
\citep{2004ApJ...603L..89K}.  It had been expected that frame-dragging
effects dominate those of oblateness
\citep{1998ApJ...509L..37K,1999ApJ...513..827M}. However, we now show
for the first time that the vertical epicyclic frequency
 in prograde circular orbits around a neutron star may be even
larger than the orbital frequency ($\omega_z-\Omega_{\rm K}>0$).
  As this occurs for
astrophysically interesting masses, the effect could have important
consequences for models of QPOs (see Section ~\ref{disc}), some of
which have already been studied for quark stars
\citep{2014PhRvD..89j4001G}.

Figure~1 shows the scaled orbital and epicyclic frequencies versus
gravitational mass (in units of mass of the Sun)  at $r=1.3a$ for
uniformly rotating neutron stars (left panel) and strange quark stars
(right panel) rotating at a fixed frequency $f_{\rm rot} = 600$~Hz.
While at the astrophysically expected masses of $M>M_\odot$ the
vertical epicyclic frequency is less than the orbital one
($\omega_{z}<\Omega_K$)
for the neutron star modeled with the FPS equation of state
(as expected for a metric close to the Kerr one),
for lower masses the opposite relation holds,
$\omega_{z}>\Omega_K$, as a result of stellar oblateness.
 However, for higher masses effects of strong gravity, such as frame
dragging, dominate the qualitative behavior of the frequency curves.
For the quark star model, the results are qualitatively similar,
except that the change in sign of $\omega_{z}-\Omega_K$
takes place at a higher mass value ($M\approx 1.1M_\odot$ for the quark star
versus $M\approx 0.9M_\odot$ for the FPS neutron star).

Figure~2 shows the same effects for more rapidly rotating stars
 ($f_{\rm rot} =900\,$Hz).
The frequencies are presented for $r=a$, i.e., at the stellar equator.
Interestingly, the effects of oblateness on the epicyclic frequencies
are now seen to qualitatively affect the ordering of the frequencies
at typical pulsar masses
($\omega_z>\Omega_K$ for the FPS neutron stars with values of $M$ up to
the canonical mass of $1.4M_\odot$, while for quark stars this is true
up to  $1.7M_\odot$).  In part this is because of the higher
rotation rate for these models, i.e., their larger oblateness,
and in part because the orbits are closer
to the star in Fig.~2 than in Fig.~1---the higher multipoles decay rapidly
 with the radial distance,
so their effect is more pronounced near the star.

\begin{figure}
\vspace{5mm}
\begin{center}
\includegraphics[width=0.45\linewidth,height=.35\linewidth]{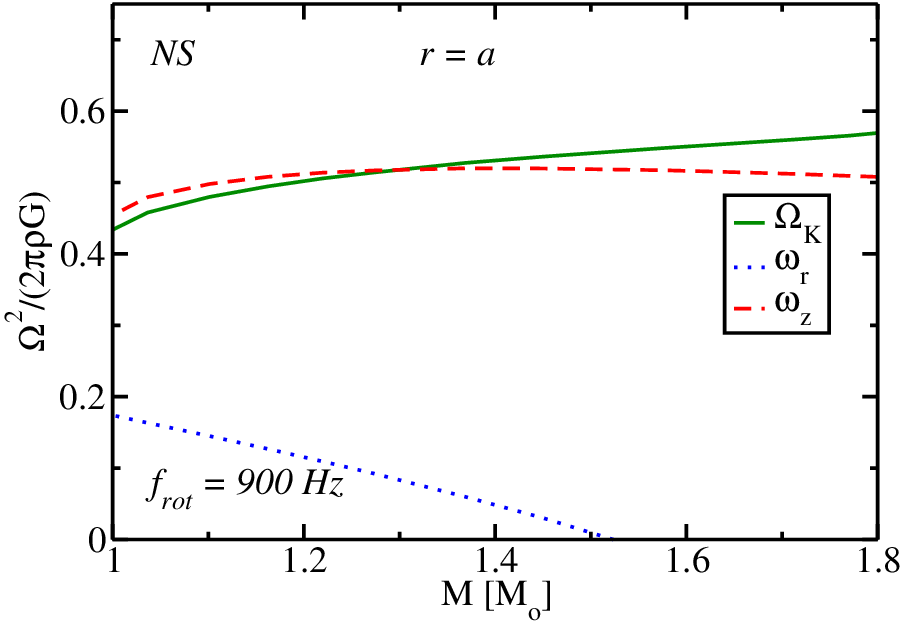}
\hspace{5mm}
\includegraphics[width=0.45\linewidth,height=.35\linewidth]{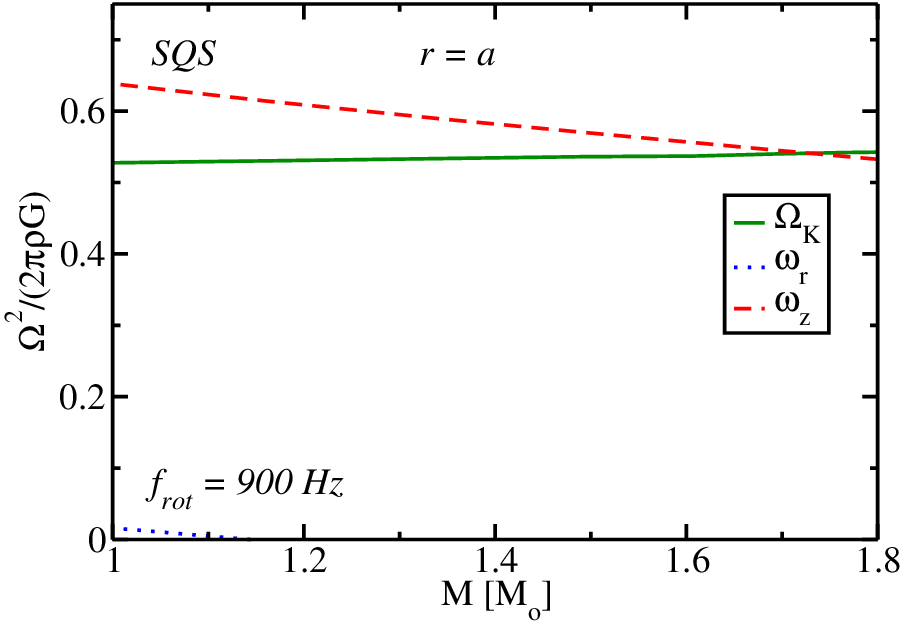}
\end{center}
\caption{Squares of the orbital and epicyclic frequencies (scaled with 2$\pi G\rho$) at the stellar equator ($r = a$) versus gravitational mass for uniformly rotating neutron stars (left panel), and strange quark stars (right panel), rotating at a fixed frequency $f_{\rm rot}$ = 900~Hz. The solid (green) line corresponds to orbital frequency, the dashed (red) line to vertical epicyclic frequency and the dotted (blue) line to radial epicyclic frequency.}
\end{figure}

\section{Discussion and conclusions}
\label{disc}

The purpose of this work is to study in GR the influence of oblateness on
the orbital and epicyclic frequencies for rapidly rotating neutron
stars and to compare the results to analogous results for strange
quark stars \citep{2014PhRvD..89j4001G}. Surprisingly, we have found
that  effects of oblateness familiar from Newtonian studies
\citep{2013MNRAS.434.2825K}, such as decreasing of the radial
epicyclic frequency with the stellar rotation rate and the vertical
epicyclic frequency exceeding the orbital one ($\omega_{z}>\Omega_K$),
are present  for realistic models of rotating neutron stars in
general relativity.

Epicyclic frequencies determine the properties of oscillation modes of
thin accretion disks \citep{1980PASJ...32..377K,1999PhR...311..259W}.
One of the most promising modes that may correspond to the observed
QPOs, the $c$-mode, is described by a corrugation of the disk
precessing  at a frequency close to $\Omega_K-\omega_{z}$, and it may
be present only if $\omega_{z}<\Omega_K$ \citep{2001ApJ...548..335S}.
Our results indicate that for some neutron stars (at least for the FPS
equation of state) the latter condition may not hold throughout the
inner accretion disk. This could indicate the necessity of revisiting
the QPO models.  In another class of models
\citep[e.g.,][]{2008NewAR..51..841K}, one of the kHz QPOs could
correspond directly to motion with the frequency $\omega_{z}$. One
possibility that could now be taken into account is that the higher of
the twin kHz QPO frequencies may have a value larger then the orbital
frequency ($\omega_{z}>\Omega_K$).

\ack
This work was supported in part by the POMOST/2012-6/11 Program of
Foundation for Polish Science  co-financed by the European Union within the
European Regional Development Fund, by the MAESTRO grant 2013/08/A/ST9/00795
of the Polish NCN, and the NewCompStar COST Action.

\bibliography{pmateuszf}
\end{document}